# Dependence of chaotic behavior on optical properties and electrostatic effects in double beam torsional Casimir actuation

F. Tajik[1,4], M. Sedighi[2], A.A. Masoudi[1], H. Waalkens[3] and G. Palasantzas[4]

[1]Department of Physics, Alzahra University, Tehran 1993891167, Iran

[2]New Technologies Research Center (NTRC), Amirkabir University of Technology, Tehran 15875-4413, Iran

[3]Bernoulli Institute for Mathematics, Computer Science and Artificial Intelligence, University of Groningen, Nijenborgh 9, 9747 AG Groningen, The Netherlands

[4]Zernike Institute for Advanced Materials, University of Groningen, Nijenborgh 4, 9747 AG Groningen, The Netherlands

**Abstract**

We investigate the influence of Casimir and electrostatic torques on double beam torsional microelectromechanical systems with materials covering a broad range of conductivities of more than three orders of magnitude. For the frictionless autonomous systems, bifurcation and phase space analysis shows that there is a significant difference between stable and unstable operating regimes for equal and unequal applied voltages on both sides of the double torsional system giving rise to heteroclinic and homoclinic orbits, respectively. For equal applied voltages, only the position of a symmetric unstable saddle equilibrium point is dependent on the material optical properties and electrostatic effects, while in any other case there are stable and unstable equilibrium points are dependent on both factors. For the periodically driven system, a Melnikov function approach is used to show the presence of chaotic motion rendering predictions of whether stiction or stable actuation will take place over long times impossible. Chaotic behavior introduces significant risk for stiction, and it is more prominent to occur for the more conductive systems that experience stronger Casimir forces and torques. Indeed, when unequal voltages are applied, the sensitive dependence of chaotic motion on electrostatics is more pronounced for the highest conductivity systems.

PACS numbers: 64.70.Nd, 85.85.+j, 12.20.Fv

# I. Introduction

Current advancement in fabrication of microelectromechanical systems (MEMS) and subsequent dimension miniaturization towards nanoelectromechanical systems (NEMS) warrant careful consideration of Casimir forces in the analysis and design of these systems [1-3]. The Casimir force can have a significant magnitude in these systems because of the relatively large surface areas and small gaps between mechanical elements, which under certain conditions can undergo jump-to-contact and permanent adhesion which is a phenomenon known as stiction [4]. As a matter of fact, the Casimir forces between two objects arise due to perturbation of quantum fluctuations of the electromagnetic (EM) field, as it was predicted by H. B. Casimir in 1948 [5] assuming two perfectly reflecting parallel plates. E. M. Lifshitz and coworkers in the 50's [6] considered the general case of real dielectric plates by exploiting the fluctuation-dissipation theorem which relates the dissipative properties of the plates due to optical absorption by many microscopic dipoles and the resulting EM fluctuations. This theory describes the attractive interaction due to quantum fluctuations at all separations covering both the Casimir (long-range) and van der Waals (short-range) regimes [1-9].

The dependence of the Casimir force on material properties is an important topic because, in principle, one can tailor the force by engineering the boundary conditions of the electromagnetic field with a suitable choice of materials. This allows the exploration of new concepts in actuation dynamics in devices via the control of the magnitude of the Casimir force and torque using different materials with a variety of optical properties [10-20]. Although the electrostatic force can, in principle, be switched off, by letting the applied potential tend to zero, the Casimir force will always be present even at absolute zero temperature and can influence the actuation dynamics of micro/nano devices [3,9-11]. So far, several studies have been performed to investigate the Casimir torque in torsional actuators, which arise due to broken rotational symmetry [21-23] or misalignment between two optical axes [24-27], in addition to mechanically driven torques due to the normal Casimir force [11, 28-32]. Indeed, torsional actuators find applications to torsional radio frequency (RF) switches, tunable torsional capacitors, and torsional micro mirrors [1-3,7,8]. They are composed of two electrodes, where one is fixed and the other can rotate freely around an axis toward the fixed one when a voltage is applied [9]. A very useful configuration is the double beam torsional actuator, which is also used in high precision Casimir force measurements for table top laboratory cosmology [7, 28].

However, detailed exploration of the chaotic dynamics of the double beam system with respect to stiction phenomena, and for different interacting materials, is still missing. By shrinking the size of these devices an unavoidable problem could be the occurrence of chaotic motion leading to device malfunction. This phenomenon causes abrupt change in the dynamical behavior and eventually leads to stiction hampering long term device predictability [11, 33, 34]. Therefore, in this paper we will investigate the actuation dynamic of a torsional double beam actuating under the influence of electrostatic and Casimir torques, with electrodes made from materials with a wide and diverse range of optical properties including gold (Au), phase change materials (PCMs), and conductive silicon carbide (SiC). This is especially important for the double beam configuration since the phase space analysis shows increased complexity due to switching between heteroclinic and homoclinic orbits in absence and presence of electrostatic balancing forces, respectively.

## II. Materials and double beam actuation system

In order to cover a wide range of materials with different optical properties we have chosen Au as a good metal conductor [7, 8], the crystalline (C) state of the PCM AIST ($Ag_5In_5Sb_{60}Te_{30}$) [15, 16] as an intermediate conductivity system, and nitrogen doped SiC as a poor conductor though a suitable material for operation in harsh environments [20, 35]. Note that the PCMs are renowned for their use in optical data storage (Blue-Rays, DVDs etc.), where AIST in particular during switching between the amorphous and crystalline phases yields a ~20-25 % Casimir force contrast at separations ~100 nm [15, 16]. Indeed, for comparison the static conductivity ratio $\omega_p^2/\omega_\tau$ in terms of the Drude model, with $\omega_p$ the plasma frequency and $\omega_\tau$ the relaxation frequency, gives for these materials $\omega_p^2/\omega_\tau|_{SiC} = 0.4$ eV for SiC [20], $\omega_p^2/\omega_\tau|_{AIST(C)} = 10.1$ eV for AIST [16], and $\omega_p^2/\omega_\tau|_{Au} \approx = 1600$ eV for Au [18]. These values indicate a conductivity contrast with respect to Au, which is a very good conductor, of more than three orders of magnitude. It should be noted that for the less conductive systems, e.g. SiC, we assume sufficiently thick coatings to ignore the contribution on the Casimir force of the underline basic material (e.g. Si) that is used for the fabrication of the beams in MEMS. In addition, the chosen materials show significant optical contrast for the dielectric function at imaginary frequencies $\varepsilon(i\xi)$, which is a necessary input for the Casimir force calculations via the Lifshitz theory, for frequencies $\xi<1$ eV and will manifest in Casimir force variations for nanoscale separations $c/2\xi>10$ nm (see Fig. 1 and the Appendix for the extrapolations of measured optical data).

The equation of motion for the double beam torsional system (inset Fig. 1), where the fixed plate is considered to be coated by Au and the rotating plate by another conducting material of choice (e.g. Au, SiC, and AIST), is given by

$$I_0 \frac{d^2\theta}{dt^2} + I_0 \frac{\omega}{Q} \frac{d\theta}{dt} = \tau_{res} + \tau_{elec} + \tau_{Cas} \qquad (1)$$

with $I_0$ the moment of rotation inertia. The mechanical Casimir torque $\tau_{Cas}$ is given by [36]

$$\tau_{Cas} = \int_0^{L_x} r(F_{Cas}^R(d_R') - F_{Cas}^L(d_L'))L_y \, dr, \qquad (2)$$

where $F_{Cas}^{R,L}(d_{R,L}')$ is the Casimir force that is calculated using Lifshitz theory (see the Appendix ). $L_x'$ (=$2L_x$) and $L_y$ are the length and width, respectively, of each plate (where we consider $L_x = L_y = 10\mu m$). $F_{Cas}^R(d_R')$ and $F_{Cas}^L(d_L')$ refer to the Casimir force on the right and left part of the rotating plate, with $d_R' = d - L_x \sin(\theta)$ and $d_L' = d + L_x \sin(\theta)$, respectively. The initial distance when the plates are parallel is assumed to be d=200 nm, and the system temperature is fixed at T=300 K.

The total effective electrostatic torque $\tau_{elec}$ acting on the rotating plate is given by $\tau_{elec} = \tau_{elec}^R - \tau_{elec}^L$, where $\tau_{elec}^R$ and $\tau_{elec}^L$ are the electrostatic torques due to the applied potentials $V_a^R$ and $V_a^L$ at the right and left end of the rotating plate, respectively. Upon substitution of the torques $\tau_{elec}^{R,\,L}$ [31, 36, 37] we obtain

$$\tau_{elec} = \frac{1}{2}\varepsilon_0 L_y (V_a^R - V_c)^2 \frac{1}{\sin^2(\theta)} \left[ \ln\left(\frac{d_R'}{d}\right) + \frac{L_x \sin(\theta)}{d_R'} \right]$$

$$-\frac{1}{2}\varepsilon_0 L_y (V_a^L - V_c)^2 \frac{1}{\sin^2(\theta)} \left[ \ln\left(\frac{d_L'}{d}\right) - \frac{L_x \sin(\theta)}{d_L'} \right]. \qquad (3)$$

In Eq. (3) $\varepsilon_0$ is the permittivity of vacuum, and $V_c$ is the contact potential difference between the interacting materials of the plates [15]. For simplicity, we will consider only the potential difference $V_{L,R} = V_a^{L,R} - V_c$ for the torque calculations.

Finally, in Eq. (1) both the Casimir and electrostatic torques are counterbalanced by the restoring torque $\tau_{res} = -k\theta$ with k the torsional spring constant at the support point of the rotating beam [38]. The term $I_0(\omega/Q)(d\theta/dt)$ in Eq.(1) is the intrinsic energy dissipation of the moving beam with Q the quality factor. Initially, we will consider high quality factors $Q \geq 10^4$ [39, 40] and neglect the effect of dissipation. The frequency $\omega$ is assumed to have a value that is typical for many resonators like AFM cantilevers, and MEMS [7, 28, 39, 40]. Notably the type of motion we consider here applies when the beam does not elastically deform because we assume large beam lengths ($L_x$) and small torsional angles at maximum separation ($\theta_0 = d/L_x = 0.02 \ll 1$).

### III. Results and discussion

In order to investigate the effect of optical properties on the actuation of the torsional double beam, we introduce the bifurcation parameter $\delta_{Cas} = \tau_{Cas}^M / k\theta_0$ that represents the ratio of the maximal Casimir torque $\tau_{Cas}^M = \tau_{Cas}(\theta = \theta_0)$ (for the Au-Au system) to the maximum restoring torque $k\theta_0$. $\delta_{Cas}$ will help us to determine when there is a stable periodic solution for the torsional system that corresponds to sufficient restoring torque to prevent stiction of the plates [41, 42]. Using $\delta_{Cas}$, Eq. 1 assumes the more convenient form

$$\frac{d^2\varphi}{dT^2} + \varepsilon \frac{1}{Q}\frac{d\varphi}{dT} = -\varphi + \delta_v \frac{1}{\varphi^2}\left[\ln(1-\varphi) + \frac{\varphi}{1-\varphi} - p^2\left[\ln(1+\varphi) - \frac{\varphi}{1+\varphi}\right]\right]$$

$$+ \delta_{Cas}\left[\frac{\tau_{cas}}{\tau_{Cas}^M}\right] + \varepsilon \frac{\tau_0}{\tau_{res}^{Max}}\cos\left(\frac{\omega}{\omega_0}T\right) \quad (4)$$

with $\varphi = \theta/\theta_0$, $T = \omega_0 t$, $I = I_0/k$ and p voltage ratio $p = V_L/V_R$. $\delta_v = (\varepsilon_0 V_R^2 L_y L_x^3)/(2kd^3)$ is the corresponding electrostatic bifurcation parameter [11, 43]. The parameter $\varepsilon$ was introduced to distinguish between the conservative frictionless and autonomous operation of the torsional system ($\varepsilon = 0$), and the non-conservative operation with friction and an additional external periodic driving term ($\varepsilon = 1$).

**(a) Conservative system ($\varepsilon=0$)**

The conservative system is the starting point of our stability analysis of the torsional system. The equilibrium points for conservative motion are obtained by the condition $\tau_{total} = \tau_{res} + \tau_{elec} + \tau_{Cas} = 0$. The latter yields from Eq. (4)

$$-\varphi + \delta_v \frac{1}{\varphi^2}\left[\ln(1-\varphi) + \frac{\varphi}{1-\varphi} - p^2[\ln(1+\varphi) - \frac{\varphi}{1+\varphi}]\right] + \delta_{Cas}\left[\frac{\tau_{cas}}{\tau_{Cas}^M}\right] = 0. \quad (5)$$

Figure 2 shows plots of $\delta_{Cas}$ vs. $\varphi$ for all studied materials with and without applied voltage. For double beam torsional MEMS, there is a significant difference between the Casimir bifurcation curves when there is electrostatic balance (no applied voltage with $\delta_v = 0$, or similarly $V_R=V_L \neq 0$ with p=1), in comparison with the unbalanced case when a voltage is applied to only one end of the beam (e.g., $V_R>0$ and $V_L=0$ and equivalently p=0) or both ends have a voltage but with different magnitude ($V_R \neq V_L$ and p≠1). In fact, when the electrostatic torque has equal magnitude at both ends of the beam, then the equilibirium points shown in the bifurcation diagram (except for $\varphi = 0$) are always unstable. Obviously, when the electrostatic potential is applied on one end of the beam (p=0 and $V_R>0$) then the system shows the bifurcation diagrams of a single torsional beam [11].

Let us elaborate on the similarities and differences of the Casimir bifurcation diagrams. In fact, in both the balanced and the unbalanced case the bifurcation diagrams differ especially near the maximum. The systems approach critical unstable and stable behavior for the unbalanced and balanced cases, respectively, in an order that is determined by the magnitude of their conductivity (from Au, AIST(C), and SiC). In Fig. 2(b), which belongs to the unbalanced situation, the solid lines show stable regions where the restoring torque is strong enough to produce stable motion (since $\delta_{Cas} \sim 1/k$). Notably, in Fig. 2 the dashed lines indicate unstable regions, where the torsional MEM loses its stability, and stiction occurs for motion close to the fixed plate. Two coexisting equilibrium points occur in the unbalanced situation for $\delta_{Cas} < \delta_{Cas}^{MAX}$. The equilibrium point closer to $\varphi = 0$ (solid line) is stable, and the other one close to $\varphi = 1$ (dashed line) is unstable. In the unbalanced case however, when $\delta_{Cas} < \delta_{Cas}^{MAX}$ the bifurcation curves show only one unstable equilibrium for the system while there is always a stable equilibrium point at $\varphi = 0$. The unstable equilibria satisfy the additional condition d $\tau$ total/ $d\varphi = 0$, which yields

$$-1 + \delta_v \left[ \frac{2\varphi - 3}{\varphi^2(1-\varphi)^2} + \frac{2\ln(1-\varphi)}{\varphi^3} - P^2 \left[ \frac{2\varphi+3}{\varphi^2(1+\varphi)^2} + \frac{2\ln(1-\varphi)}{\varphi^3} \right] \right] + \delta_{Cas} \frac{1}{\tau_{Cas}^m} \left( \frac{d\tau_{Cas}}{d\varphi} \right) = 0. \quad (6)$$

By increasing $\delta_{Cas}$ or weakening the restoring torque since $\delta_{Cas} \sim 1/k$, the distance between the stable and unstable points becomes smaller until one reaches the maximum point $\delta_{Cas}^{MAX}$ (for both balanced and unbalanced cases) which satisfies both Eqs.(5) and (6). According to Fig. (2) when $\delta_{Cas} \approx \delta_{Cas}^{MAX}$ for the Au-Au system, then for the other actuating systems, which have a lower conductivity and experience a less strong Casimir torque, it is still the condition $\delta_{Cas} < \delta_{Cas}^{MAX}$ indicating an increased range for stable motion. In other words, with decreasing restoring torque or equivalently decreasing spring constant k, the Au-Au torsional device will lose sooner its stability region in comparison to the other interacting systems.

The insets in Figs. 2(b) and 2(c) depict the sensitivity of the stable and unstable regions in torsional MEMS for both optical properties and electrostatics. Indeed, if the applied voltage increases then $\delta_{Cas}^{MAX}$ decreases for all systems. Due to the attractive nature of the electrostatic force, the device would require a higher restoring torque to preserve the stable operation of the system during motion. The range of the torsional angles that covers the stable region also decreases by increasing voltage. The dependence of the electrostatic bifurcation parameter $\delta_v$ on optical properties and electrostatics is shown in Figs. 3 and 4. Not only the maximum $\delta_v^{MAX}$ decreases similar to $\delta_{Cas}^{MAX}$, but also the range of the torsional angles (distance between stable center and unstable saddle point) becomes shorter by increasing material conductivity (increasing $\delta_{Cas}$) and/or applied voltage. Therefore, the range of bifurcation parameters to produce periodic motion ($0 < \delta_{Cas} < \delta_{Cas}^{MAX}$ and $\delta_v > 0$) is increased with decreasing material conductivity in MEMS. Note that for $\delta_{Cas} > \delta_{Cas}^{MAX}$ there is no stability in the torsional device even in absence of electrostatic torque ($\delta_v = 0$).

Besides the bifurcation diagrams, the phase space portraits also show the sensitive dependence of the actuation dynamics on optical properties and electrostatics [44, 33, 11]. Figure 5 shows the phase portrait for the Au-Au system for both the absence and presence of applied voltage. Clearly, for the balanced case, there is a heteroclinic orbit that separates unstable motion from the stable closed orbits around the stable center point. Indeed, for the balanced situation ($\delta_v = 0$ or $p = 1$), the stable center of the symmetric torsional device is at $\varphi = 0$ but there are two symmetric unstable equilibrium points of opposite sign. The stable central equilibrium point is completely independent of optical properties and the magnitude of the applied voltage for p=1. By contrast, as Fig. 5(b) shows, for the unbalanced situation, the inequality of electrostatics yields a homoclinic orbit that separates stable and unstable motion. Here the stable and unstable equilibrium points do not show any symmetry in the phase portrait, and both of them are strongly dependent on the optical properties and applied voltages [11].

For later comparison with the driven case ($\varepsilon = 1$), we use plots of the transient time to collapse of the moving beam on the ground plate referred to as stiction. These are shown in Figs. 6 and 7. For both the balanced and the unbalanced case, the size of the area enclosed by the heteroclinic and homoclinic orbits, respectively, decreases when the conductivity of the interacting materials increases. For any initial conditions in the region outside the region enclosed by the heteroclinic and homoclinic orbits, the moving beam will perform unstable motion and quickly collapse on the ground plate (except for the small set of points contained in the stable curves of the saddles). The reduction in size of the enclosed area by decreasing conductivity confirms again that torsional systems with higher conductivity materials lose their stability sooner because of the existence of stronger Casimir forces and torques between the plates. Moreover, an increase of an applied unbalanced voltage can have a stronger influence on the reduction of the area enclosed by the homoclinic orbit as compared to the area enclosed by the heteroclinic orbit in the balanced case.

### (b) Periodically driven system ($\varepsilon = 1$)

We furthermore performed calculations to investigate the existence of chaotic behavior of the torsional system when undergoing forced oscillation due to a time periodic applied external torque $\tau_0 \cos(\omega t)$ [33]. Chaotic behavior occurs if the separatrix resulting from the heteroclininc or homoclinic orbits of the conservative system split and have transversal intersections. The occurrence of transversal intersections can be inferred from the presence of zeroes of the so-called Melnikov [30, 35]. If we $\varphi_{het}^C(T)$ and $\varphi_{hom}^C(T)$ denote the heteroclinic and homoclinic solutions, respectively, of the conservative system then the Melnikov functions for the torsional system are given by [11, 33, 34, 44]

$$M^{het}(T_0) = \frac{1}{Q}\int_{-\infty}^{+\infty}(\frac{d\varphi_{het}^C(T)}{dT})^2 \, dT + \frac{\tau_0}{\tau_{res}^{MAX}}\int_{-\infty}^{+\infty}\frac{d\varphi_{het}^C(T)}{dT} \cos[\frac{\omega}{\omega_0}(T-T_0)] \, dT, \qquad (7)$$

and

$$M^{hom}(T_0) = \frac{1}{Q}\int_{-\infty}^{+\infty}(\frac{d\varphi_{hom}^C(T)}{dT})^2 \, dT + \frac{\tau_0}{\tau_{res}^{MAX}}\int_{-\infty}^{+\infty}\frac{d\varphi_{hom}^C(T)}{dT} \cos[\frac{\omega}{\omega_0}(T+T_0)] \, dT. \qquad (8)$$

The separatrix splits and as a consequence chaotic motion occurs if the Melnikov function has simple zeros, i.e. $M^{het/hom}(T_0) = 0$ and $(M^{het/hom})'(T_0) \neq 0$. Equality in the latter condition corresponds to the limiting case of a double zero and gives the threshold condition for the occurrence of chaotic motion [33, 44]. If we define

$$\mu_{het/hom}^{C} = \int_{-\infty}^{+\infty} \left(\frac{d\varphi_{het/hom}^{C}(T)}{dT}\right)^2 dT, \quad \beta_{het/hom}(\omega) = \left|H\left[\text{Re}\left(F\left\{\frac{d\varphi_{het/hom}^{C}(T)}{dT}\right\}\right)\right]\right|, \quad (9)$$

and $\alpha = (1/Q)(\tau_0/\tau_{res}^{MAX})^{-1} = \gamma\omega_{0\,\theta_0}/\tau_0$ where $\gamma = I\omega_o/Q$, and $H[...]$ denotes the Hilbert transform (see [33, 44]) then the threshold condition for chaotic motion becomes

$$\alpha = \beta_{het/hom}(\omega)/\mu_{het/hom}^{C}. \quad (10)$$

Figure 8 shows the threshold curves $\alpha = \gamma\omega_{0\theta_0}/\tau_0$ vs. the driving frequency ratio $\omega/\omega_o$. For large values of $\alpha$ (above the curve), the dissipation dominates the driving torque leading to regular oscillatory motion near the stable equilibrium point of the conservative system. However, for parameter values below the curve, the transversal intersections of stable and unstable manifolds causes chaotic motion. Clearly for systems with higher conductivity, which lead to stronger Casimir torques, chaotic motion is more likely to occur as is manifested by the larger area below the threshold curves. Figures 8(b) and 8(c) show the strong dependence of the region below the threshold curve on the applied voltage for the unbalanced and balanced cases, respectively. The presence of an electrostatic torque clearly changes the threshold curves in a more profound way for systems with higher conductivity.

Figures 9 and 10 show plots of the transient times to stiction for different values of the threshold parameter α from Fig. 8 for all materials studied here. When chaotic motion occurs for small values of α, then there is a region of initial conditions where the prediction of the behavior of the oscillating system is a highly formidable task or even impossible. If we compare with Figure 6 and 7 where chaotic motion does not occur, then there is in the presence of chaos no simple smooth boundary between the regions of stable and unstable solutions (the red and the blue regions in the figures). As a result, the jump to contact instability leading to stiction could take place after several periods affecting the long-term stability of the device. Therefore, chaotic behavior introduces significant risk for stiction, and this is more prominent to occur for the more conductive systems that experience an increasing Casimir torque. And again, for parameters related to Fig. 8(b) when unbalanced voltages are applied, Fig. 11 illustrates the sensitive dependence of chaotic motion on the applied electrostatic potential for the highest conductivity system Au-Au.

## IV. Conclusions

In conclusion, we have explored the influence of Casimir and electrostatic torques on double beam torsional microelectromechanical systems with materials covering a broad range of conductivities of more than three orders of magnitude. For the conservative systems, bifurcation and phase space analysis have shown that there is a significant difference between stable and unstable operating regimes for equal and unequal applied voltages on both sides of the double torsional system displaying heteroclinic and homoclinic orbits, respectively. For equal, applied voltages, only the position of the symmetric unstable saddle equilibrium point is dependent on the material optical properties and electrostatic effects, while in any other case both stable and unstable points are dependent on both factors. For the non-conservative system, Melnikov function and Poincare phase space analysis have shown the presence of chaotic motion making impossible to predict whether stiction or stable actuation will take place on a long term basis. Chaotic behavior introduces significant risk for stiction, and it is more prominent to occur for the more conductive systems that experience increasing Casimir forces and torques. Indeed, when unequal voltages are applied, the sensitive dependence of chaotic motion on electrostatics is more pronounced for the highest conductivity systems. Finally, our analysis can provide more insight for the design of double beam torsional systems using proper materials in order to ensure wider range of stable operation for both fundamental force measurements and technology applications of double beam type electromechanical systems.

## Acknowledgements

GP acknowledges support from the Zernike Institute of Advanced Materials, University of Groningen. M. S. acknowledges support from the Amirkabir University of Technology. FT and AAM acknowledge support from the Department of Physics at Alzahra University.

**APPENDIX: Brief Lifshitz theory and Dielectric function of materials with extrapolations**

The Casimir force $F_{Cas}(d)$ in Eq.(2) is given by [6]

$$F_{Cas}(d) = \frac{k_B T}{\pi} \sum_{l=0}^{\infty} {}' \sum_{\nu=TE,TM} \int_0^\infty dk_\perp \, k_\perp \, k_0 \, \frac{r_\nu^{(1)} r_\nu^{(2)} \exp(-2k_0 d)}{1 - r_\nu^{(1)} r_\nu^{(2)} \exp(-2k_0 d)}. \tag{A.1}$$

The prime in the first summation indicates that the term corresponding to $l = 0$ should be multiplied with a factor $1/2$. The Fresnel reflection coefficients are given by $r_{TE}^{(i)} = (k_0 - k_i)/(k_0 + k_i)$ and $r_{TM}^{(i)} = (\varepsilon_i k_0 - \varepsilon_0 k_i)/(\varepsilon_i k_0 + \varepsilon_0 k_i)$ for the transverse electric (TE) and magnetic (TM) field polarizations, respectively. $k_i = \sqrt{\varepsilon_i(i\xi_l) + k_\perp^2}$ ($i = 0,1,2$) represents the out-off plane wave vector in the gap between the interacting plates ($k_0$) and in each of the interacting plates ($k_{i=(1,2)}$). $k_\perp$ is the in-plane wave vector.

Furthermore, $\varepsilon(i\xi)$ is the dielectric function evaluated at imaginary frequencies, which is the necessary input for calculating the Casimir force between real materials using Lifshitz theory. The latter is given by [6]

$$\varepsilon(i\xi) = 1 + \frac{2}{\pi} \int_0^\infty \frac{\omega \, \varepsilon''(\omega)}{\omega^2 + \xi^2} \, d\omega. \tag{A2}$$

For the calculation of the integral in Eq. (A2) one needs the measured data for the imaginary part $\varepsilon''(\omega)$ of the frequency dependent dielectric function $\varepsilon(\omega)$. The materials were optically characterized by ellipsometry over a wide range of frequencies at J. A.Woollam Co.: VUV-VASE (0.5–9.34 eV) and IR-VASE (0.03–0.5 eV)) [16]. The experimental data for the imaginary part of the dielectric function cover only a limit range of frequencies $\omega_1$ ($= 0.03$ ev) $< \omega < \omega_2$ ($= 8.9$ ev). Therefore, for the low optical frequencies ($\omega < \omega_1$) we extrapolated using the Drude model for the crystalline phase [16]

$$\varepsilon''_L(\omega) = \frac{\omega_p^2 \, \omega_\tau}{\omega \, (\omega^2 + \omega_\tau^2)}, \tag{A3}$$

where $\omega_p$ is the plasma frequency, and $\omega_\tau$ is the relaxation frequency. Furthermore, for the high optical frequencies ($\omega > \omega_2$) we extrapolated using [16]

$$\varepsilon''_H(\omega) = \frac{A}{\omega^3}. \tag{A.4}$$

Finally, using Eqs. (A2)-(A4), the function $\varepsilon(i\xi)$ is given by

$$\varepsilon(i\xi)_C = 1 + \frac{2}{\pi} \int_{\omega_1}^{\omega_2} \frac{\omega \, \varepsilon''_{exp}(\omega)}{\omega^2 + \xi^2} \, d\omega + \Delta_L \varepsilon(i\xi) + \Delta_H \varepsilon(i\xi), \tag{A.5}$$

with

$$\Delta_L \varepsilon(i\xi) = \frac{2}{\pi} \int_0^{\omega_1} \frac{\omega \, \varepsilon''_L(\omega)}{\omega^2 + \xi^2} \, d\omega, \text{ and } \Delta_H \varepsilon(i\xi) = \frac{2}{\pi} \int_{\omega_2}^{\infty} \frac{\omega \, \varepsilon''_H(\omega)}{\omega^2 + \xi^2} \, d\omega. \tag{A.6}$$

**Figure captions**

**Figure 1** Dielectric functions at imaginary frequencies ε(iξ) for Au, SiC, and crystalline (C) AIST, which were used for the Casimir torque calculations. The inset shows the double beam torsional system.

**Figure 2** (a) Bifurcation diagrams $\delta_{Cas}$ vs. φ with (a) $\delta_v = 0$; (b) $\delta_v = 0.05$ and p=0 (the inset shows similar plots for $\delta_v = 0.4$); and (c) $\delta_v = 0.05$ and p=1 (the inset shows similar plots for $\delta_v = 0.4$). The solid and dashed lines represent the stable and unstable equilibrium points, respectively.

**Figure 3** Bifurcation diagrams $\delta_v$ vs. φ for different value of $\delta_{Cas}$ for the Au-Au system for (a) p=1, and (b) p=0.

**Figure 4** Bifurcation diagrams $\delta_v$ vs. φ for $\delta_{Cas}$ =750 for all studied materials (a) p=1, and (b) p=0.

**Figure 5** Phase portraits dφ/dt vs. φ for $\delta_{Cas} = 250$ of the Au-Au torsional system and initial conditions inside and outside of (a) the heteroclinic orbit with $\delta_v = 0$ (balanced system), (b) the homoclinic orbit with $\delta_v = 0.23$ and p=0 (unbalanced system), and (c) the heteroclinic orbit with $\delta_v = 0.23$ and p=1 (balanced system).

**Figure 6** Contour plot of the transient time to stiction for initial conditions in the $\varphi$ - $d\varphi/dt$ phase plane for the conservative system with: right column $\delta_{Cas} = 750$ and $\delta_v = 0$; and left column $\delta_{Cas} = 750$, $\delta_v = 0.07$, and p=1. For the calculations, we used 150×150 initial conditions ($\varphi$, $d\varphi/dt$). The red region contains initial conditions that lead to stable oscillations. The heteroclinic orbits (right column) and homoclinic orbit (left column) separates sharply stable and unstable solutions reflecting the absence of chaotic behavior.

**Figure 7** Contour plot of the transient time to stiction for initial conditions in the $\varphi$ - $d\varphi/dt$ phase plane for the conservative system with: right column $\delta_{Cas} = 750$, $\delta_v = 0.02$, and p=0.5; and left column $\delta_{Cas} = 750$, $\delta_v = 0.02$, and p=0. For the calculations we used 150×150 initial conditions ($\varphi$, $d\varphi/dt$). The red region contains initial conditions for which the torsional device is performing stable oscillations. The heteroclinic orbits (right column) and homoclinic orbit (left column) separates sharply stable and unstable solutions reflecting the absence of chaotic behavior.

**Figure 8** Threshold curve $\alpha \left(= \gamma \omega_{0\theta_0}/\tau_0\right)$ vs. driving frequency $\omega/\omega_o$ (with $\omega_o$ the natural frequency of the system). The area bellow the curve corresponds to parameters that can lead to chaotic motion with $\delta_{Cas} = 750$: (a) $\delta_v = 0$, (b) $\delta_v = 0.02$ with p=0, and (c) $\delta_v = 0.07$ with p=1.

**Figure 9** Contour plot of the transient times to stiction using Poincare phase maps $d\varphi/dt$ vs. $\varphi$ for the non-conservative system with: right column $\delta_{Cas} = 750$, $\delta_v = 0.07$, p=1 (balanced situation) and $\alpha = 0.8$; left column $\delta_{Cas} = 750$, $\delta_v = 0.07$, p=1 (balanced situation) and $\alpha = 2$. For the calculations we used 150×150 initial conditions ($\varphi$, $d\varphi/dt$). The red region shows that initial condition for which the torsional device shows still stable motion after 100 oscillations. With decreasing $\alpha$ the chaotic behavior increases, and the area of stable motion (red region) shrinks more for the systems with higher conductivity.

**Figure 10** Contour plot of the transient times to stiction in the $\varphi$ - $d\varphi/dt$ phase plane for the non-conservative system with: right column $\delta_{Cas} = 750$, $\delta_v = 0.02$, p=0 (unbalanced situation) and $\alpha = 0.5$; and left column $\delta_{Cas} = 750$, $\delta_v = 0.02$, p=0 (unbalanced situation) and $\alpha = 4$. For the calculations, we used 150×150 initial conditions ($\varphi$, $d\varphi/dt$). The red region shows that initial condition for which the torsional device shows still stable motion after 100 oscillations. With decreasing $\alpha$ the chaotic behavior increases, and the area of stable motion (red region) shrinks more strongly for the systems with a higher conductivity.

**Figure 11** Contour plot of the transient times to stiction in the $\varphi$ - $d\varphi/dt$ phase plane for the non-conservative system with: left column $\delta_{Cas} = 750$, $\delta_v = 0$, and $\alpha = 0.5$; and right column $\delta_{Cas} = 750$, $\delta_v = 0.02$, $\alpha = 0.5$ and p=0. The systems material considered here are Au-Au and Au-SiC. For the calculations, we used 150×150 initial conditions ($\varphi$, $d\varphi/dt$). The red region shows the initial conditions for which the torsional device shows stable motion after 100 oscillations with oscillating frequency $\omega/\omega_0 = 0.2$. With increasing $\delta_v$ (or equivalently applied

voltage) the chaotic behavior increases, and the area of stable motion shrinks more strongly for the systems with a higher conductivity and applied potential.

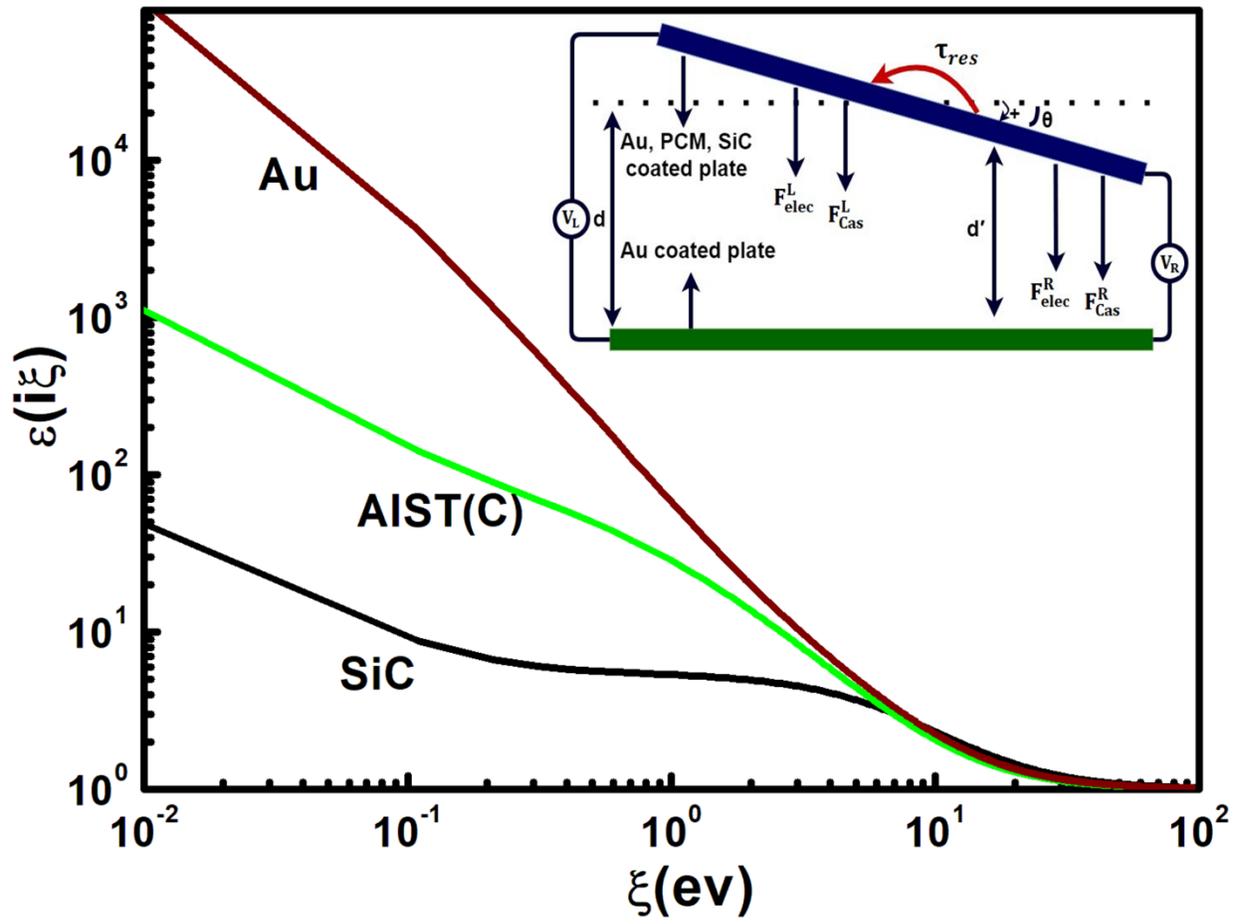

**Figure 1**

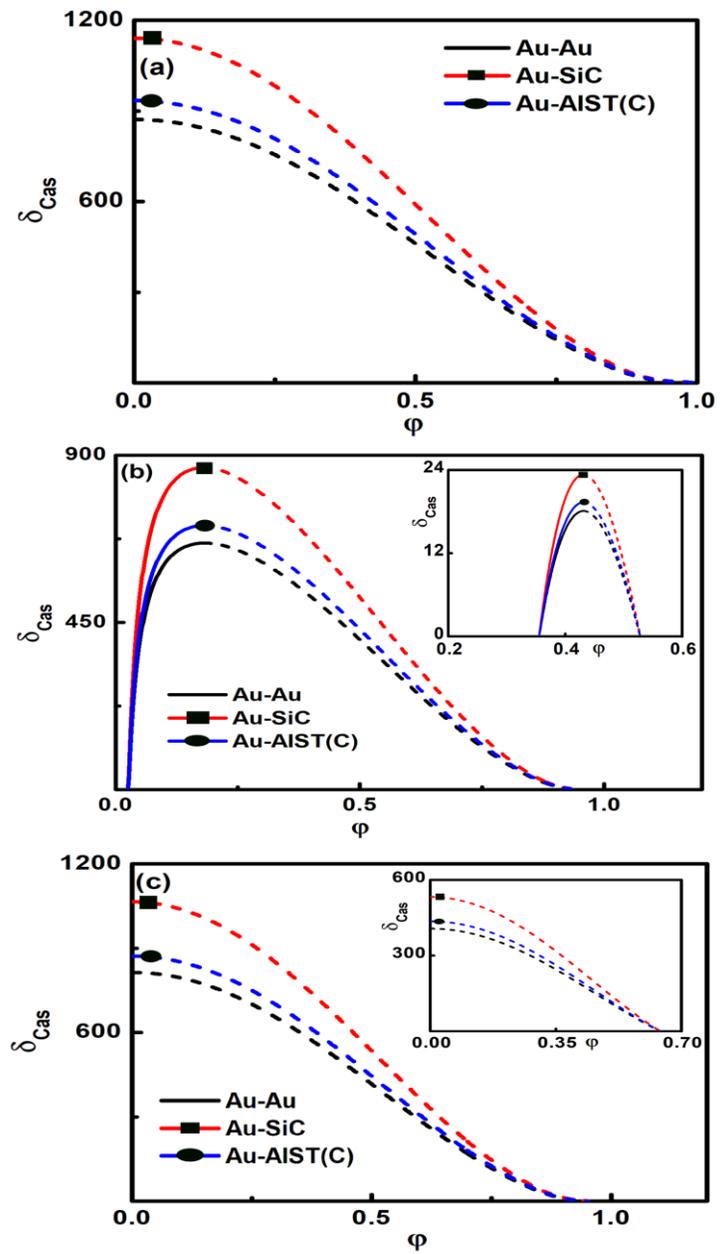

**Figure 2**

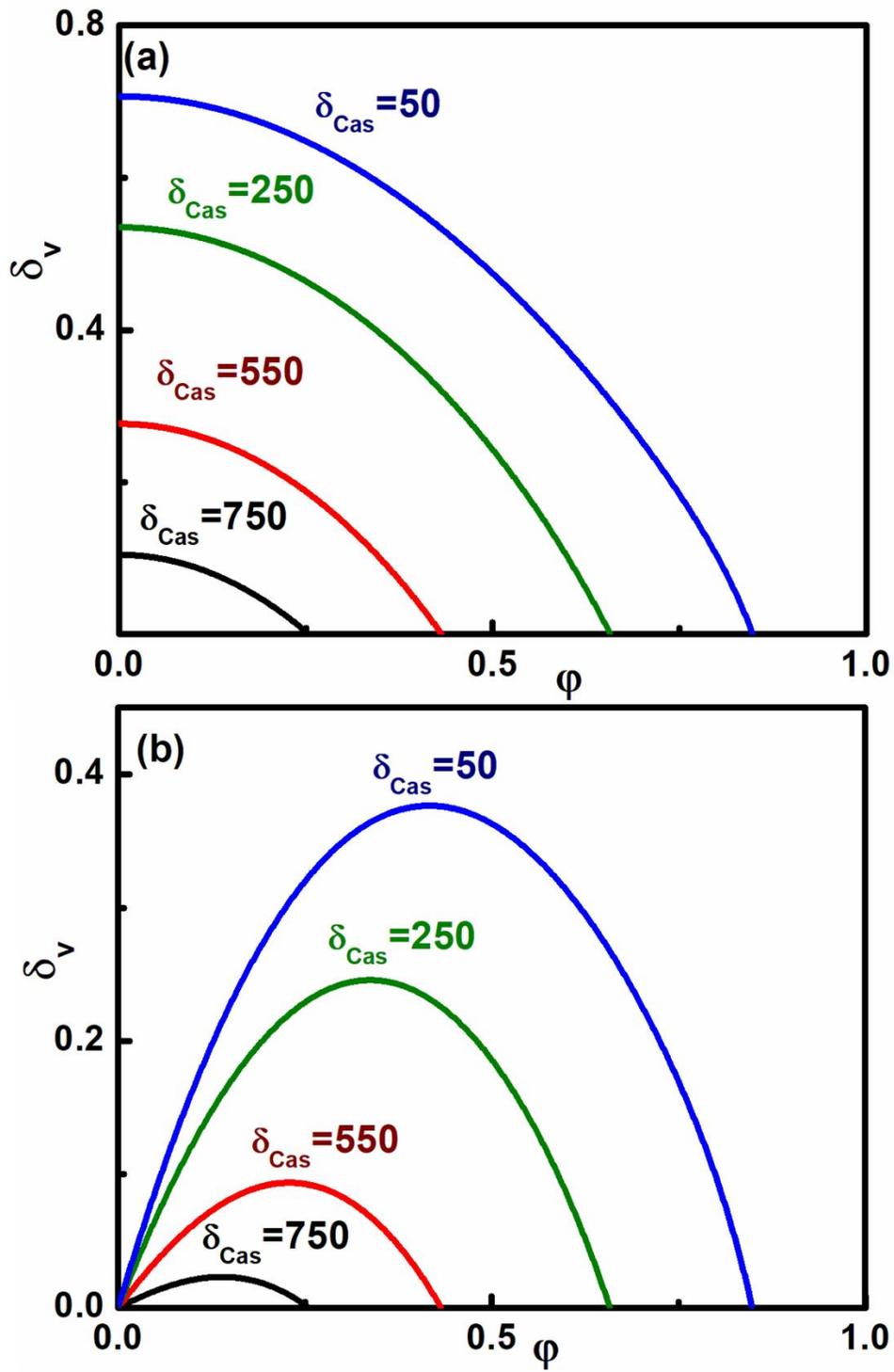

**Figure 3**

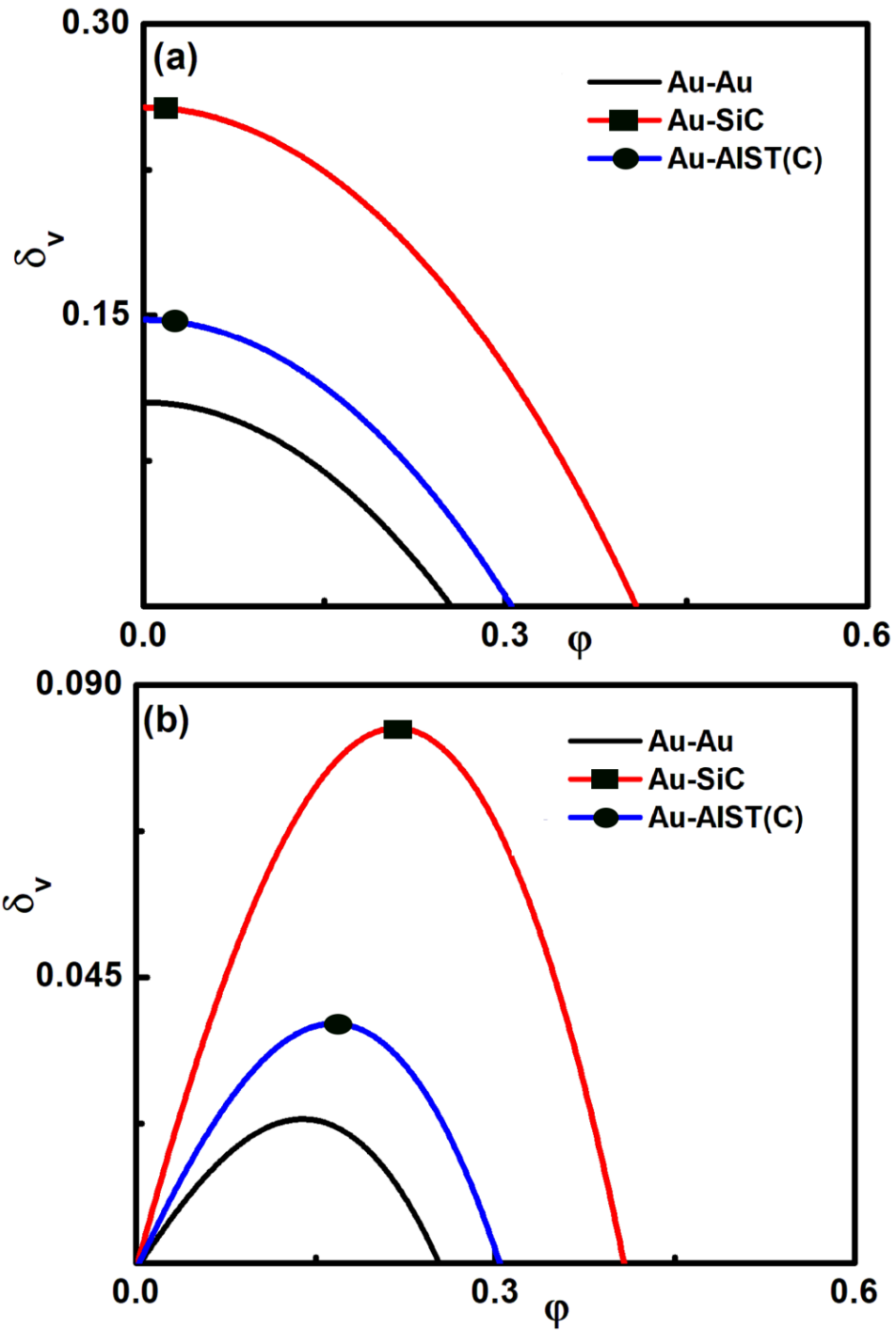

**Figure 4**

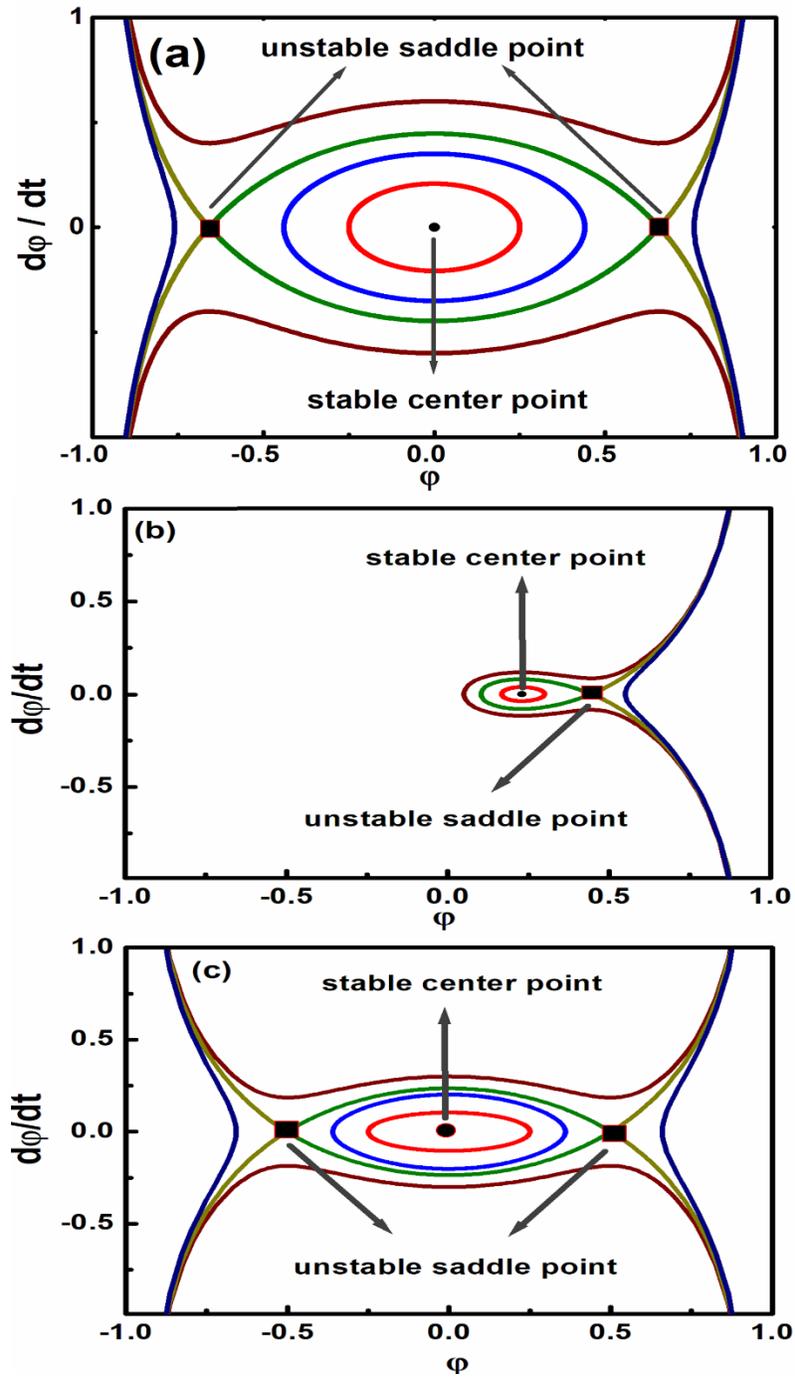

**Figure 5**

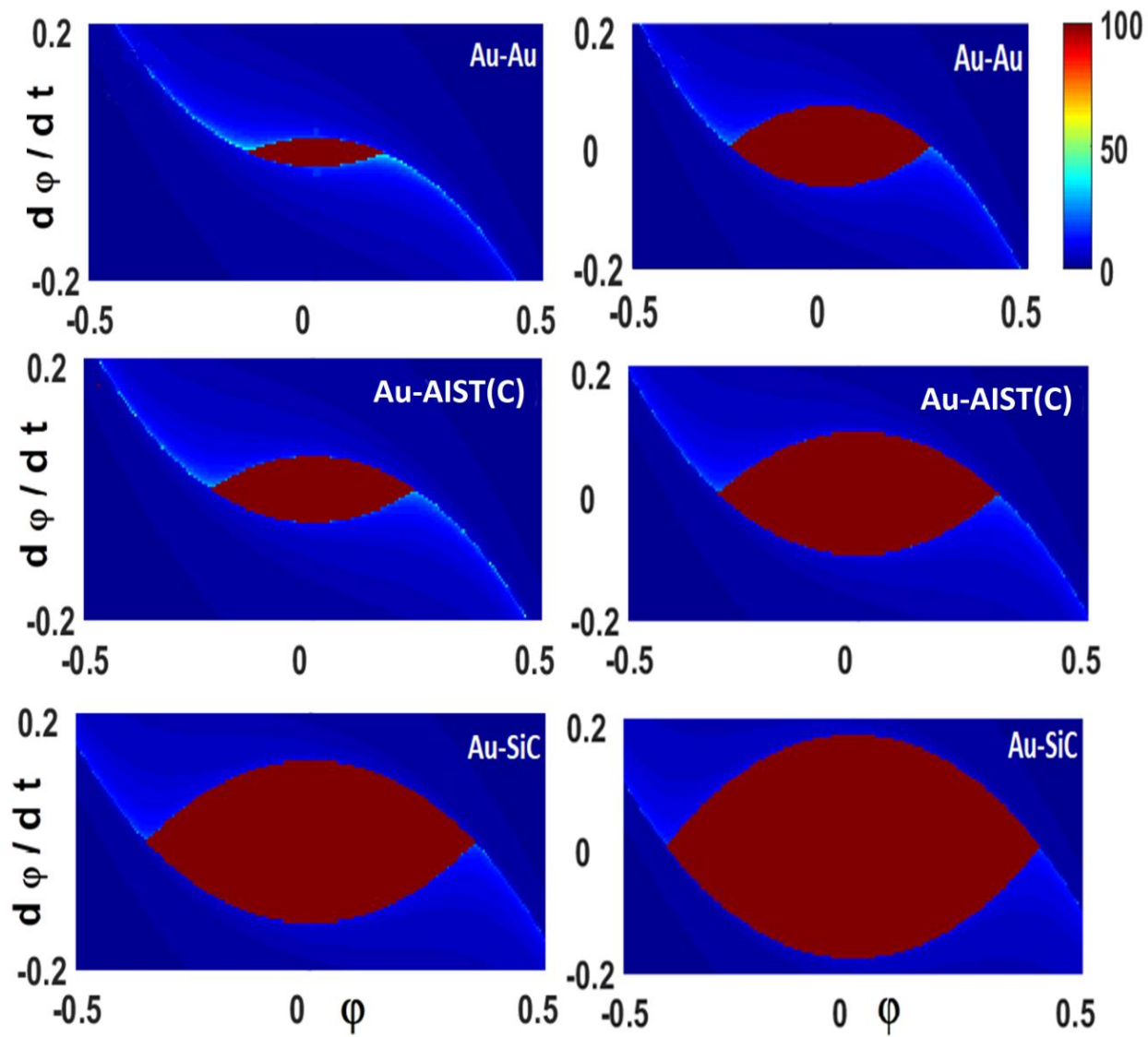

**Figure 6**

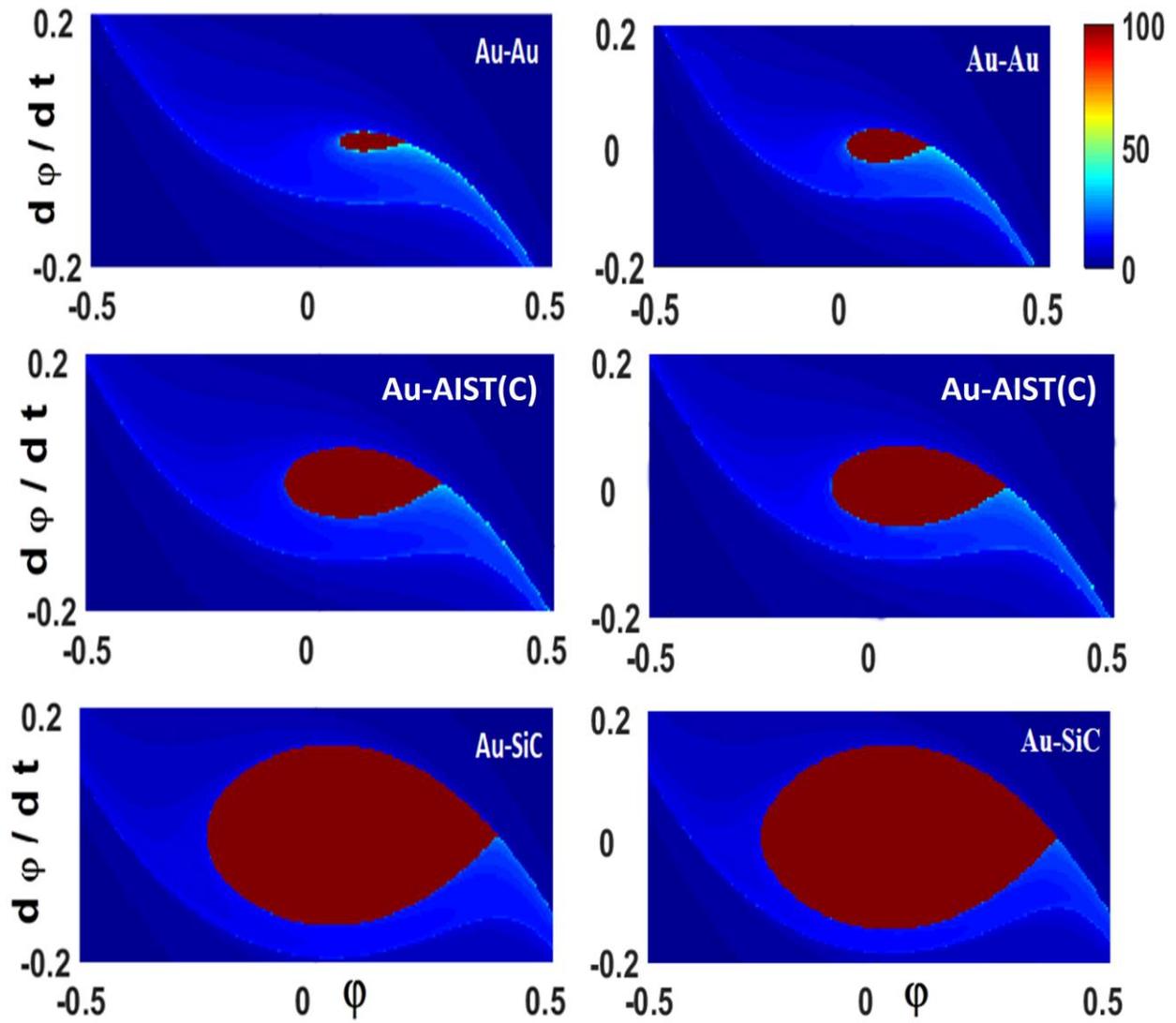

**Figure 7**

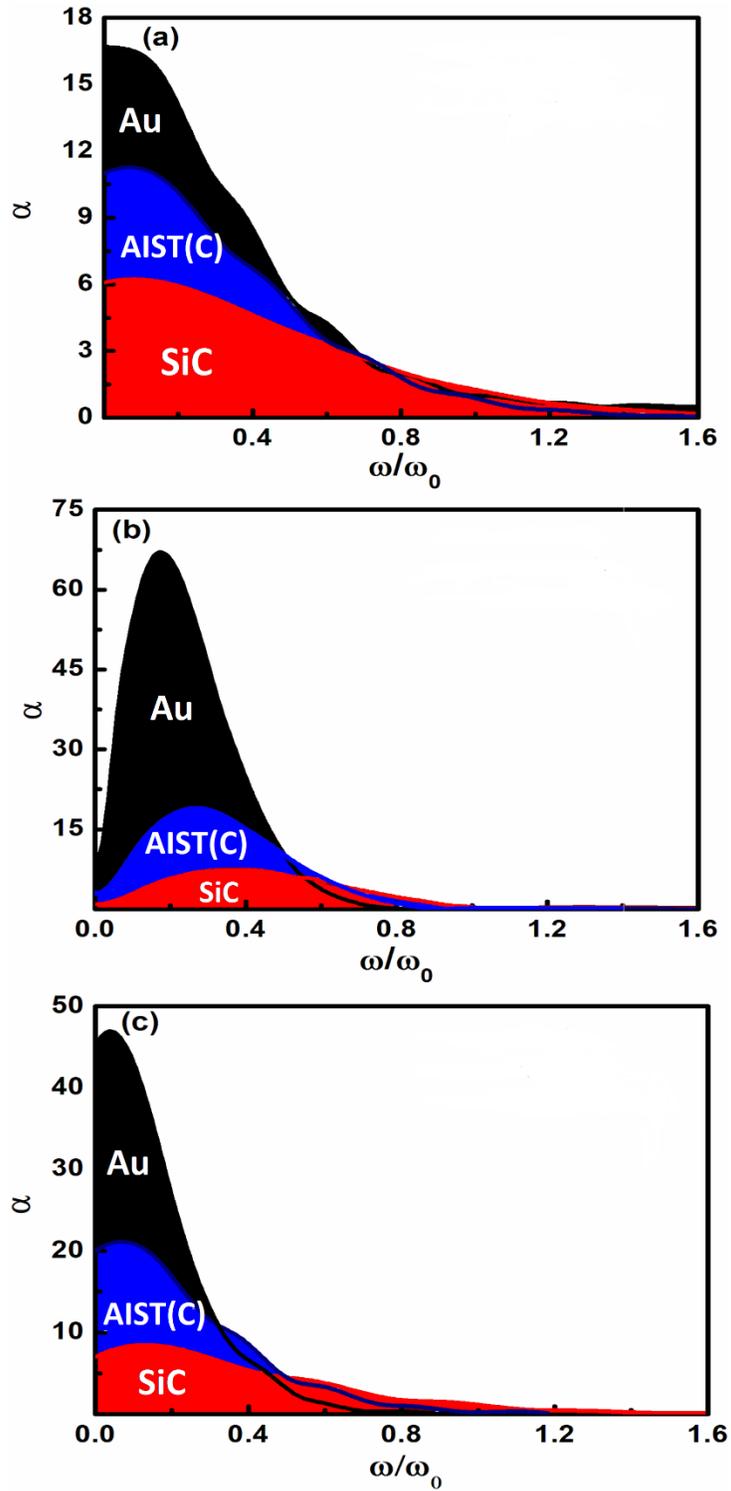

**Figure 8**

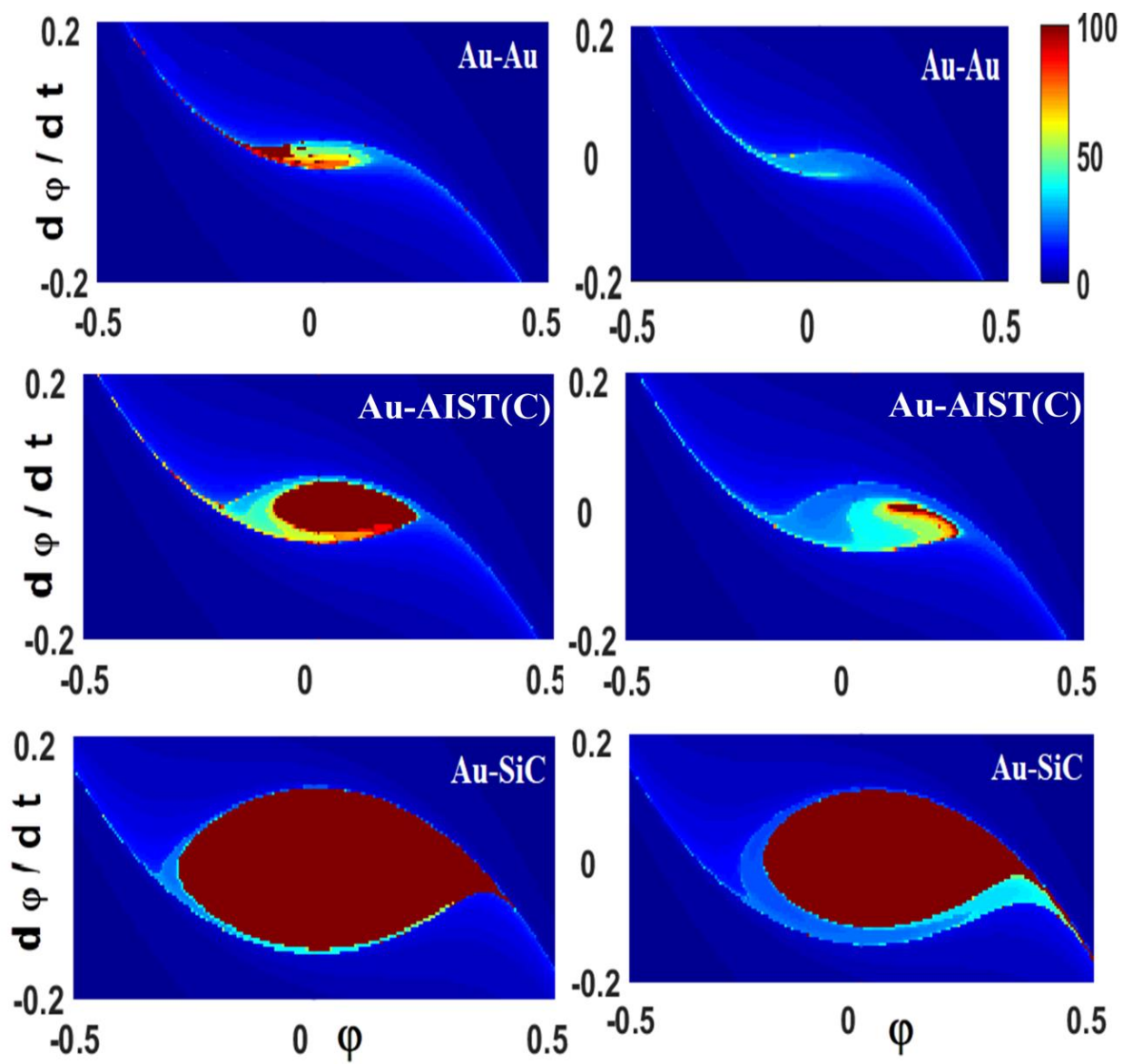

**Figure 9**

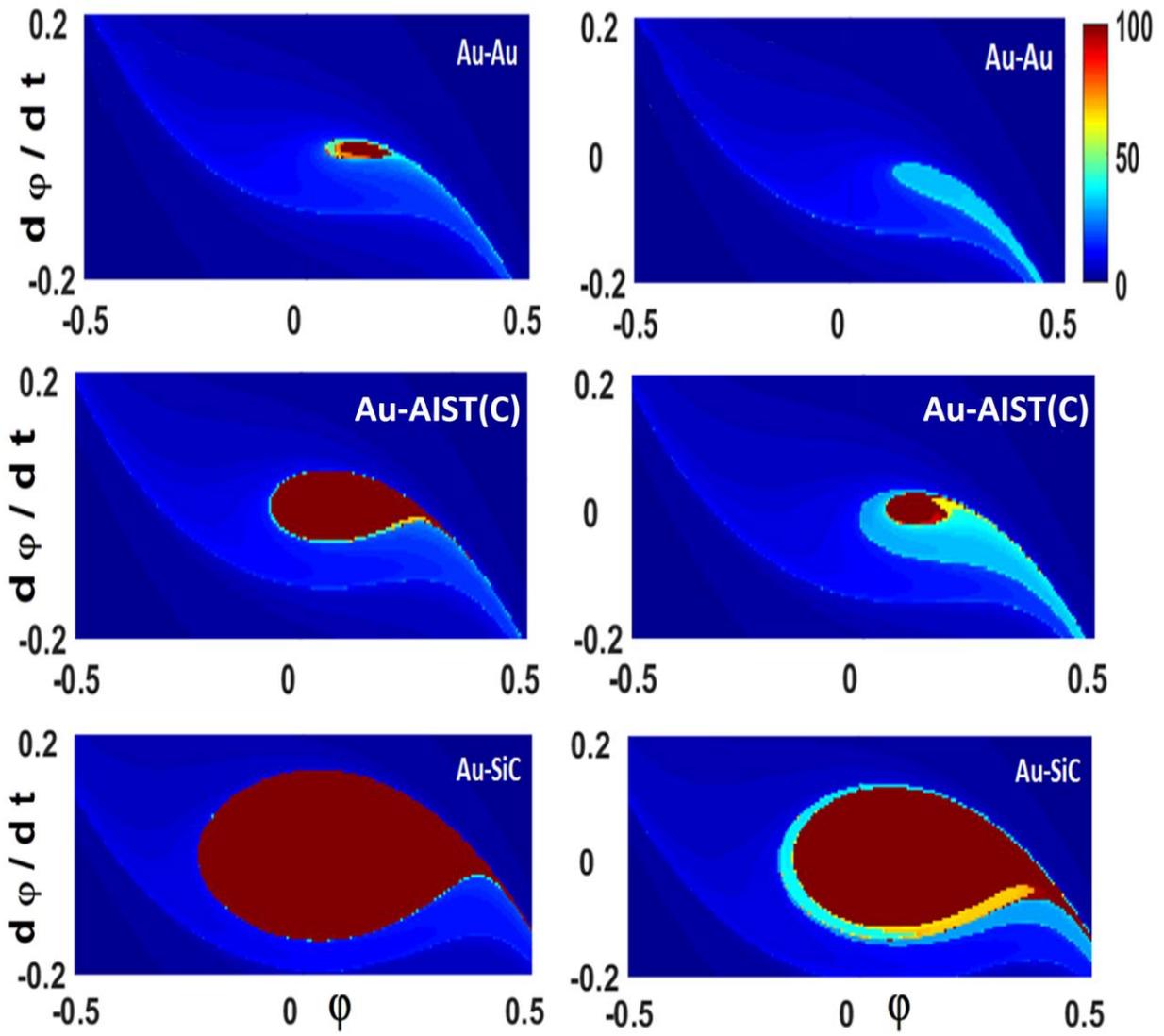

**Figure 10**

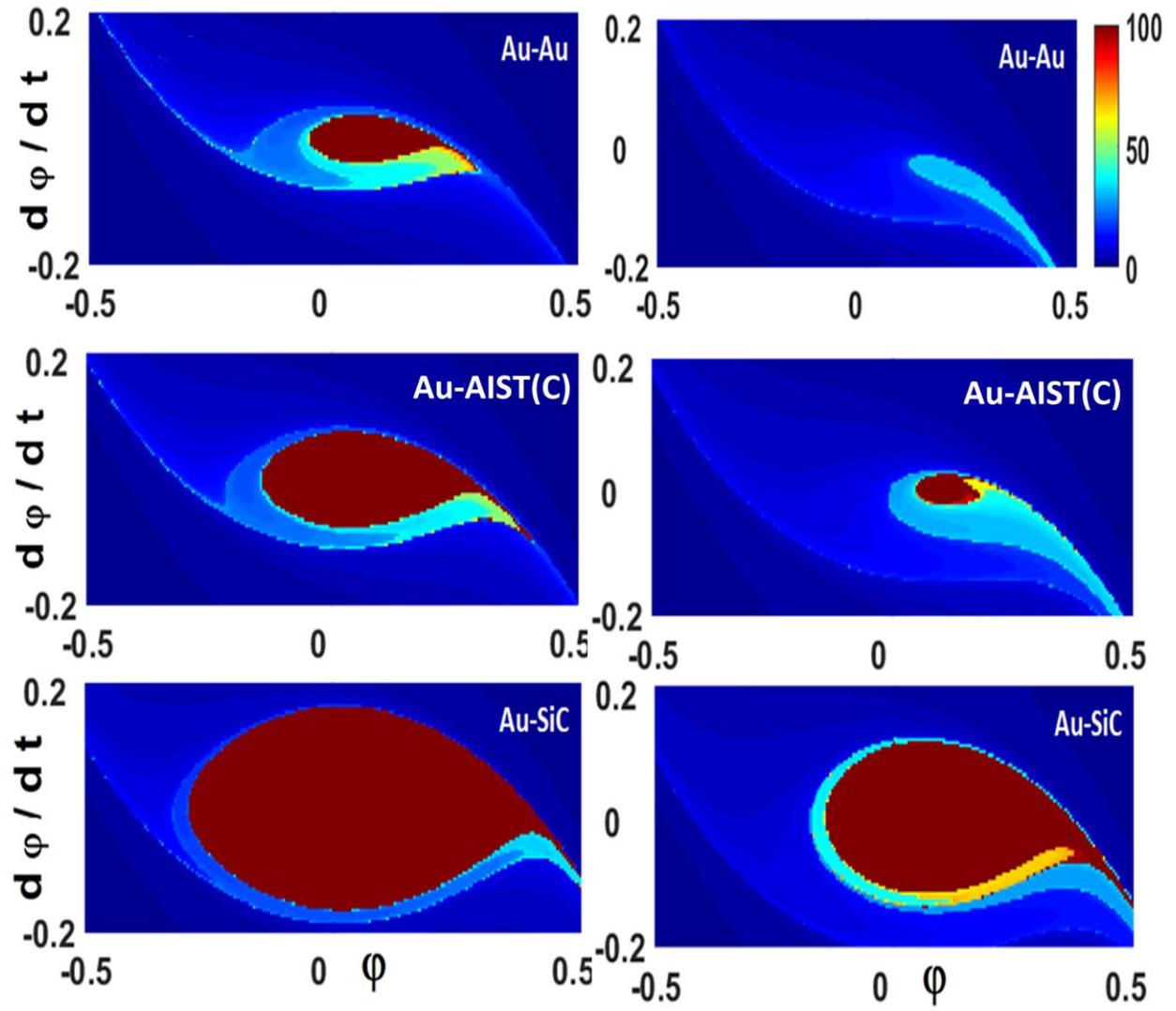

**Figure 11**